\documentclass[letterpaper,11pt]{article}
\usepackage{amsmath,amssymb,mathtools,cite,graphicx,fancyhdr}
\usepackage{bbold,color,bm,caption,authblk}
\newcommand{\dd}{\text{d}}

\usepackage{hyperref}
\hypersetup{
	unicode=false,          
	pdftitle={Optical analogue of the Schwarzschild-Planck metric},    
	pdfauthor={Alhan Moreno-Ruiz and David Bermudez},     
	pdfsubject={Analogue Gravity},   
	pdfcreator={Alhan Moreno-Ruiz and David Bermudez},   
	pdfproducer={Alhan Moreno-Ruiz and David Bermudez}, 
	pdfkeywords={theory of instabilities, analogue gravity, Hawking radiation, Bogoliubov-De Gennes equation, black hole laser, quantum, physics, mathematics}, 
	pdfnewwindow=true,      
	colorlinks=true,       	
	linkcolor=magenta,      
	citecolor=blue, 			
	filecolor=red,      		
	urlcolor=red           	
}

\begin{document}
	\lhead{Optical analogue of the Schwarzschild-Planck metric}
	\rhead{A. Moreno, D. Bermudez}
	
\title{Optical analogue of the Schwarzschild-Planck metric}
\author{Alhan Moreno\footnote{{\it email:} alhan.moreno@cinvestav.mx}, David Bermudez\footnote{{\it email:} david.bermudez@cinvestav.mx}}
\affil{\textit{Departamento de F\'{\i}sica, Cinvestav, A.P. 14-740, 07000 Ciudad de M\'exico, Mexico}}

\date{}
	
\maketitle

\begin{abstract}
	We revisit the connection between trajectories of accelerated mirrors and spacetime metrics. We present the general (1+1)D effective metric that can be obtained with a fibre-optical analogue through the Kerr effect. Then we introduce a new connection between accelerated mirrors and the optical metric. In particular, we connect them for two specific trajectories: The first one is the black mirror that perfectly recreates the Schwarzchild spacetime. The second one is the Schwarzschild-Planck metric that is a regularized version of the Schwarzschild case. The regularization depends on a length scale that has a clear physical interpretation in the fibre-optical analogue system. We study the geometric properties and the Hawking radiation produced in these new analogue metrics.
\end{abstract}

\section{Introduction}\label{sec:Introduction}

In 1916, K. Schwarzschild found the first exact solution to the Einstein equations in general relativity \cite{Schwarzschild1916}. This solution hinted at a mysterious object from which nothing can escape, not even light; these objects were later called black holes and their frontiers event horizons. In 1974, S. Hawking \cite{Hawking1974, Hawking1975} considered a semi-classical approximation: A quantum field around the classical background of the Schwarzschild metric. Under this approximation, Hawking showed that black holes radiate as black bodies with thermal spectrum and that their temperature is proportional to the surface gravity at the event horizon.

The fact that a Schwarzschild black hole radiates presents a new problem, since the predicted number of particles and radiated energy is infinite \cite{Hawking1974, Hawking1975}. Also, the Hawking mechanism probes distances smaller than the Planck length in the vicinity of the event horizon, where we expect a breakdown of the semi-classical approximation. This is the so-called trans-Planckian problem \cite{Brout.1995}. Finally, there is no observational evidence of Hawking radiation and, in fact, it seems very unlikely to obtain this evidence in the near future due to the faintness of the expected signal in comparison with the background noise from the cosmic microwave background (CMB) \cite{Noterdaeme.2011}.

In  1981, W. Unruh proposed to study the Hawking effect in analogue systems \cite{unruh.1981}. Although initially this work past almost unnoticed, its ideas took relevance in the next decade, ultimately leading to a new area of research called analogue gravity. Here, phenomena usually related to gravity are studied in different physical systems. The Hawking process in analogues leads to the production of thermal radiation that can be of classical (stimulated) \cite{Drori2019} or quantum (spontaneous) \cite{deNova2019} origin. The most common analogue systems to study Hawking radiation are water tanks \cite{Rousseaux2008}, Bose-Einstein condensates (BEC's) \cite{Barcelo.2001,Steinhauer2016}, optical fibres \cite{philbin.2008,bermudez.2016}, microcavity polaritons \cite{Nguyen.2015}, superconducting circuits \cite{Nation.2012},  and fluids of light \cite{Jacquet.2020}. 

In this work, we will focus on optical analogues. The first idea to create an optical analogue was using slow light \cite{Leonhardt2002}, but it was soon shown to suffer a number of problems while simultaneously clarifying the roles of the phase and group velocities \cite{unruh.2003}. Then in 2008, Philbin et al. \cite{philbin.2008} proposed that a localized pulse that propagates through an optical fibre recreates the horizon: the fibre-optical analogue of the event horizon.

On the other hand, S.A. Fulling and P.C.W. Davis found that an accelerated mirror propagating through empty space radiates due to the Unruh effect \cite{Fulling1976, Unruh1976, Davies1977}. Eventually, M. Good proposed a mirror trajectory that perfectly recreates the radiation from a Schwarzschild black hole, even the previously mentioned problems with infinite quantities, this is the so-called black mirror \cite{Good.2016prd, Good.2016}. The study of the connection between accelerated mirrors and spacetimes in general relativity produces new analytical tools to understand both systems. For instance, from accelerated mirrors it is easy to find the analytical Bogoliubov $\beta$ coefficient, the spectrum, the particle number, and the radiated energy \cite{Good.2016prd, Good.2013, good.2020prd, Good.2020mpl}.

Recently, M. Good and E. Linder proposed a new trajectory of an accelerated mirror that solves the problems of the Schwarzschild metric by softening the trajectories using a new length scale: The quantum-pure black mirror \cite{Good.2020mpl,Good.2019}. The corresponding spacetime metric was quickly obtained as a regularized form of the Schwarzschild spacetime called the Schwarzschild-Planck metric \cite{good.2021}.

In most cases, analogue systems only have one temporal and one spatial dimensions or (1+1)D, as is the case for fibre optics \cite{philbin.2008,bermudez.2016} and BEC's \cite{Barcelo.2001,Steinhauer2016}. Mirror trajectories consist in position as a function of time \cite{Fulling1976,Davies1977}, also regularly in (1+1)D. Then it is valuable to review the concepts of spacetime geometry in (1+1)D. In general relativity, (1+1)D systems are considered toy models that are helpful to introduce the geometry and mathematics of the theory \cite{Mann1991,Boozer.2008}. Here they can also be used to understand the geometry of the new analogue systems.   

The aim of this work is to implement the Schwarzschild-Planck metric in an optical analogue. In Section \ref{sec:G(1+1)}, we shortly revisit gravity in $(1+1)$D. In Section \ref{sec:SPmetric} we work out the mirror trajectories proposed by M. Good and collaborators and how to match the collapse condition for a shell of matter. In Section \ref{sec:Analog}, we develop the theory to recreate the general effective metric in a fibre-optical analogue. We obtain the optical analogue of the Schwarzschild-Planck metric in Section \ref{sec:Spo} and study its properties in Section \ref{sec:Properties}. Finally, we write our conclusions in Section \ref{sec:Conclusions}.

\section{Gravity in (1+1)D}\label{sec:G(1+1)}

In 1915, A. Einstein proposed the field equations that describe the interaction between matter and spacetime: the general theory of relativity \cite{einstein.1916}. A year later, K. Schwarzschild found their first exact solution with the use of several symmetries \cite{Schwarzschild1916}, but in general solving these equations is complex and exhausting. One common strategy of physicists in these difficult cases is to simplify the problem by reducing its dimensions. Here, the simplest systems have one temporal and one spatial dimensions or (1+1)D. In 1991, R. Mann proved that this kind of systems have black-hole solutions \cite{Mann1991}. However, we will see that this change of dimensions also has unforeseen physical implications.

Any stationary system in (1+1)D with the condition $|g|=-1$ can be written in orthogonal coordinates through the Gram–Schmidt process using an independent field $\alpha(x)$, as 
\begin{equation}
	\mathrm{d}s^2=\alpha(x)c^2\mathrm{d}t_0^2-\alpha(x)^{-1}\mathrm{d}x^2,
	\label{met1+1}
\end{equation}  
where $t_0$ is time and $x$ is position. Like in the usual (3+1)D systems, the horizon condition is $\alpha(x)=0$ \cite{Mann1991,Boozer.2008}. Apparently, there is an inaccessible region of spacetime, but it is only because these are not the best coordinates to describe the system \cite{Misner1973}. In this work, we explicitly write down $c$ to simplify the connection to the analogue case.

The event horizon looks like a singularity in this metric, but this is due to the ill choice of coordinates \cite{robertson.2012}. A change to the Painlev\'e \cite{Painleve.1921}, Gullstrand \cite{Gullstrand.1922} and Lema\^itre \cite{Lemaitre.1933} (PGL) coordinates solves this problem. Then, the metric takes the form 
\begin{equation}
	\mathrm{d}s^2=c^2\mathrm{d}t^2-\left(\mathrm{d}x+v(x)\mathrm{d} t\right)^2, \qquad v(x)=c\sqrt{1-\alpha(x)}.
	\label{mpgl1+1}
\end{equation}
In PGL coordinates, the metric is finite in all spacetime except at the singularity $x=0$, but including the horizon. Now, the spacetime can be seen as a Galilean fluid \cite{jacobson.2006,hamilton.2008}, the null geodesic equation is 
\begin{equation}
	\frac{\dd t}{\dd x}=-\left[v(x) \pm c \right]^{-1},
	\label{g}
\end{equation}
that defines a local inertial frame called the comoving frame, and the signs $+$ and $-$ represent copropagating and counterpropagating rays, respectively.

We will see that in analogue gravity, the black-hole spacetime is represented as a fluid that is swallowed up by the singularity. This connection to the fluid is explicit using PGL coordinates \cite{bermudez.2016,robertson.2012}. Next, we describe the general (1+1)D metric in Eq. \eqref{met1+1} by studying its geometric quantities.

\subsection{Geometric quantities}
In (1+1)D there are eight Christoffel symbols $\Gamma^\mu_{\nu\beta}$, but only six are non-zero 
\begin{equation}
	\Gamma^{0}_{01}=-\Gamma^{1}_{11}=\frac{1}{2}\alpha(x)^{-1}\alpha'(x), \qquad \Gamma^1_{00}=\frac{1}{2}\alpha(x)\alpha'(x),
	\label{chris}
\end{equation}
where $\alpha'(x)=d\alpha(x)/dx$. The other three non-zero components are obtained using the symmetry $\Gamma^\mu_{\nu\beta}=\Gamma^\mu_{\beta \nu}$. The Riemann curvature tensor $R^{\mu}_{\nu\beta \gamma}$ is made up of sixteen components with only four non-zero given by
\begin{equation}
	R^{1}_{001}=\alpha(x)^2 R^{0}_{101}=\frac{1}{2}\alpha(x)\alpha''(x).
	\label{Rtensor}
\end{equation}
We obtain the remaining two non-zero components using the property $R^{\mu}_{\nu\beta \gamma}=R^{\mu}_{\nu\gamma\beta }$. The non-zero components of the Ricci tensor $R_{\mu\nu}$ are 
\begin{equation}
	R_{00}=\frac{1}{2}\alpha(x)\alpha''(x), \quad R_{11}=-\frac{\alpha''(x)}{2\alpha(x)}.
	\label{rtensor}
\end{equation}
The Ricci $R$ and Kretschmann $K$ scalars are
\begin{equation}
	R=-\alpha''(x), \quad K=\left[\alpha''(x)\right]^2.
	\label{scalars}
\end{equation}
Finally, one of the most remarkable results in (1+1)D gravity is the fact that the Einstein tensor $G$ is always zero
\begin{equation}
	G_{\mu\nu}=R_{\mu\nu}-\frac{R}{2}g_{\mu\nu}=0.
	\label{G}
\end{equation} 
Then, the Einstein field equations in (1+1)D gravity do not constrain the metric tensor. The conclusions from this result are: (1) There are only vacuum solutions, i.e. the spacetime only has curvature, but no matter. (2) Any metric is solution of the Einstein equations in (1+1)D spacetime metric, in particular, some that have an event horizon \cite{Mann1991,Boozer.2008}.

\subsection{Surface gravity}
The surface gravity $\kappa$ is the acceleration caused directly by the spacetime geometry. Its value at the horizon allows us to assign a temperature to the black-hole geometry \cite{belgiorno.2011}. The explicit formula for $\kappa$ is
\begin{equation}
	\kappa^2=-\frac{c^4}{2}\left(\nabla^\nu k^\mu\right)\left(\nabla_\nu k_\mu\right),
\end{equation}
where $k^{\mu}$ is the Killing vector in suitable coordinates. This definition provide us $\kappa$ everywhere, not just at the horizon \cite{wald.1984}. This will be useful to connect with analogue systems later on. For the general $(1+1)$D metric with PGL coordinates in Eq. \eqref{mpgl1+1}, the surface gravity is given by
\begin{equation}
	\kappa(x)=\left|\frac{c^2}{2} \alpha'(x)\right|.
	\label{kappa}
\end{equation}

\section{Accelerated mirrors and spacetime metrics} \label{sec:SPmetric}
In 1976, S.A. Davis and P.C.W. Fulling proposed a simple model of a mirror accelerating in one direction to obtain insight into more complex processes such as the production of cosmological particles and the quantum mechanism of particle production from black holes \cite{Fulling1976,Davies1977}. Following this idea, M. Good proposed the black mirror, a trajectory of an accelerated mirror that mimics the collapse of a Schwarzschild black-hole. The accelerated mirror has the advantage that its Bogoliubov coefficients can be obtained analytically, and in consequence, it is easy to calculate the corresponding spectrum, energy, and number of particles \cite{Good.2016prd}.

\subsection{From Schwarzschild spacetime to black mirror}
\label{subsec:SP}
The Schwarzschild metric in usual coordinates is written as
\begin{equation}
	\dd s^2=f c^2\dd t^2-f^{-1}\dd r^2, \qquad f=1-\frac{r_s}{r},
\end{equation}
where $r_s=2GM/c^2$ is the Schwarzschild radius.
The analogy between the Schwarzschild black-hole and the black-mirror trajectory is worked out with the canonical collapse prescription for a shell of matter \cite{good.2021}. It is necessary to normalize the radial coordinate using the Regge-Wheeler tortoise radial coordinate $r^*$, and defining the null coordinates $f_b=t-r^*/c$ and $v=t+r^*/c$ for $r>r_s$. Since there is a singularity at $r=r_s$, another set of coordinates $U=T+r^*/c$ and $V=T-r^*/c$ is necessary to describe the spacetime for $r<r_s$. The canonical collapse prescription emerges from matching the outer ($f_b,v$) and inner ($U,V$) null coordinates at the horizon \cite{good.2021,fabbri.2005}. This prescription gives
\begin{equation}
	f_b(U)=U-\frac{c}{\kappa} \ln \left|\frac{\kappa U}{c} \right|,
	\label{ccp}
\end{equation}
where $\kappa=c^2/(2r_s)$ is the surface gravity at the horizon $r_s$. The connection between the Schwarzschild spacetime and the black mirror turns into a connection between the particle production in the gravitational collapse due the Hawking effect \cite{Hawking1974,Hawking1975} and the particle production of an accelerating mirror due to Fulling-Davis effect \cite{Fulling1976,Davies1977}.

The spectra of the produced particles in the black mirror is the same as in the Schwarzschild black hole. The energy of the produced particles comes from whatever accelerates the mirror (engine) and from the whatever causes the curvature in spacetime (mass). The trajectory of the black mirror in null coordinates is
\begin{equation}
	u_b(v)=v-\frac{c}{a}\ln\left(\frac{a v}{c}\right),
	\label{blackmirror}
\end{equation}
where $u_b=t-x/c$ is the retarded time as function of advanced time $v=t+x/c$ and $a$ is the acceleration of the mirror \cite{Good.2016prd,good.2021}. The connection is performed by matching the outer null coordinate $u_b$ (subindex $b$ is for black mirror) with the retarded time $f_b$, and making the mirror acceleration equal to the surface gravity of a Schwarzschild black-hole $a=\kappa$.

The problem is that the analogy between the black-mirror trajectory and the Schwarzschild black-hole spacetime is too perfect. The particle production at late times is in eternal equilibrium and the number of particles and radiated energy are infinite, like a black hole radiating forever \cite{good.2021}. However, this connection between mirrors and spacetimes is flexible enough to allow different systems, and in particular, one that solves these infinities.

\subsection{From black mirror to quantum-pure black mirror}
To solve these inconsistencies, M. Good, E. Linder, and F. Wilczek \cite{Good.2019,good.2021} proposed a new trajectory that still resembles an evaporating black hole, but the particle count and energy flux are finite. It produces quasi-thermal radiation that ends in a quantum-pure state \cite{Good.2019}. As it is known, when considering quantum effects the Schwarzschild background is not a stable solution anymore, as radiation evaporates the black hole, its mass is lost, and the metric changes. They called it the quantum-pure black mirror and its trajectory is

\begin{equation}
	u_p(v)=v-\frac{c}{a}\sinh^{-1}\left|\frac{g v}{c} \right|,
	\label{bmp}
\end{equation}
where $u_p$ is the new retarded time for the quantum-pure black mirror (subindex $p$ refers to pure). Now there is a new free parameter $g$ that is also an acceleration and it can be written, similarly to $\kappa$, in terms of a new length scale $g=c^2/(2\ell)$. In extensions of general relativity, it is usually considered that this new length $\ell$ is a scale factor that can be related to the microscopic structure of spacetime itself \cite{good.2021} and should be of the order of the Planck length
\begin{equation}
	\ell_P=\sqrt{\frac{\hbar G}{c^3}},
\end{equation}
where $\hbar$, $G$ and $c$ are the fundamental constants of quantum mechanics and general relativity. We will argue next that in our fibre-optical analogue, this parameter corresponds to the response time of the fibre molecules to the optical field.

\subsection{From quantum-pure black mirror to Schwarzschild-Planck metric}

Good et al. \cite{good.2021} considered the trajectory of a quantum-pure black mirror in Eq. \eqref{bmp} and used the collapse condition as in the black hole case. They obtained the following collapse condition that matches the outer $f_p$ and inner $U$ null coordinates:
\begin{equation}
	f_p(U)=U-\frac{c}{\kappa} \sinh^{-1}\left|\frac{U}{2c\ell} \right|.
\end{equation}
Just like in the black-hole case, the surface gravity is defined as $\kappa=c^2/(2r_s)$. Following the steps from Section \ref{subsec:SP} in reverse, we obtain the metric related to the collapse prescription. This leads to a new tortoise coordinate $\bar{r}^*$ defined by
\begin{equation}
	\bar{r}^*=r+r_\text{s} \sinh^{-1}\left| \frac{r-r_\text{s}}{\ell}\right|.
\end{equation}
And from the condition $\dd\bar{r}^*/\dd r\equiv\bar{f}^{-1}$ \cite{good.2021}, the so-called Schwarzschild-Planck metric is obtained as
\begin{equation}
	\dd s^2=\bar{f} c^2\dd t^2-\bar{f}^{-1}\dd r^2, \qquad \bar{f}=1-\frac{r_\text{s}}{r_\text{s}+\sqrt{(r-r_\text{s})^2+\ell ^2}}.
	\label{sp}
\end{equation}
The introduction of the scale factor $\ell$ regularizes the conventional Schwarzschild metric in a non-trivial way. As expected, the Schwarzschild metric is recovered in the limit $\ell\rightarrow 0$.

In PGL coordinates these metrics take the form of the equation \eqref{mpgl1+1}, with the function $v(r)$
\begin{equation}
	v(r)=c\sqrt{\frac{r_s}{r}}, \qquad v(r)= c\sqrt{\frac{r_\text{s}}{r_\text{s}+\sqrt{(r-r_\text{s})^2+\ell^2}}},
	\label{metpgl}
\end{equation}
for the Schwarzschild and Schwarzschild-Planck metric, respectively.

\begin{figure}
	\centering
	\includegraphics[height=50mm]{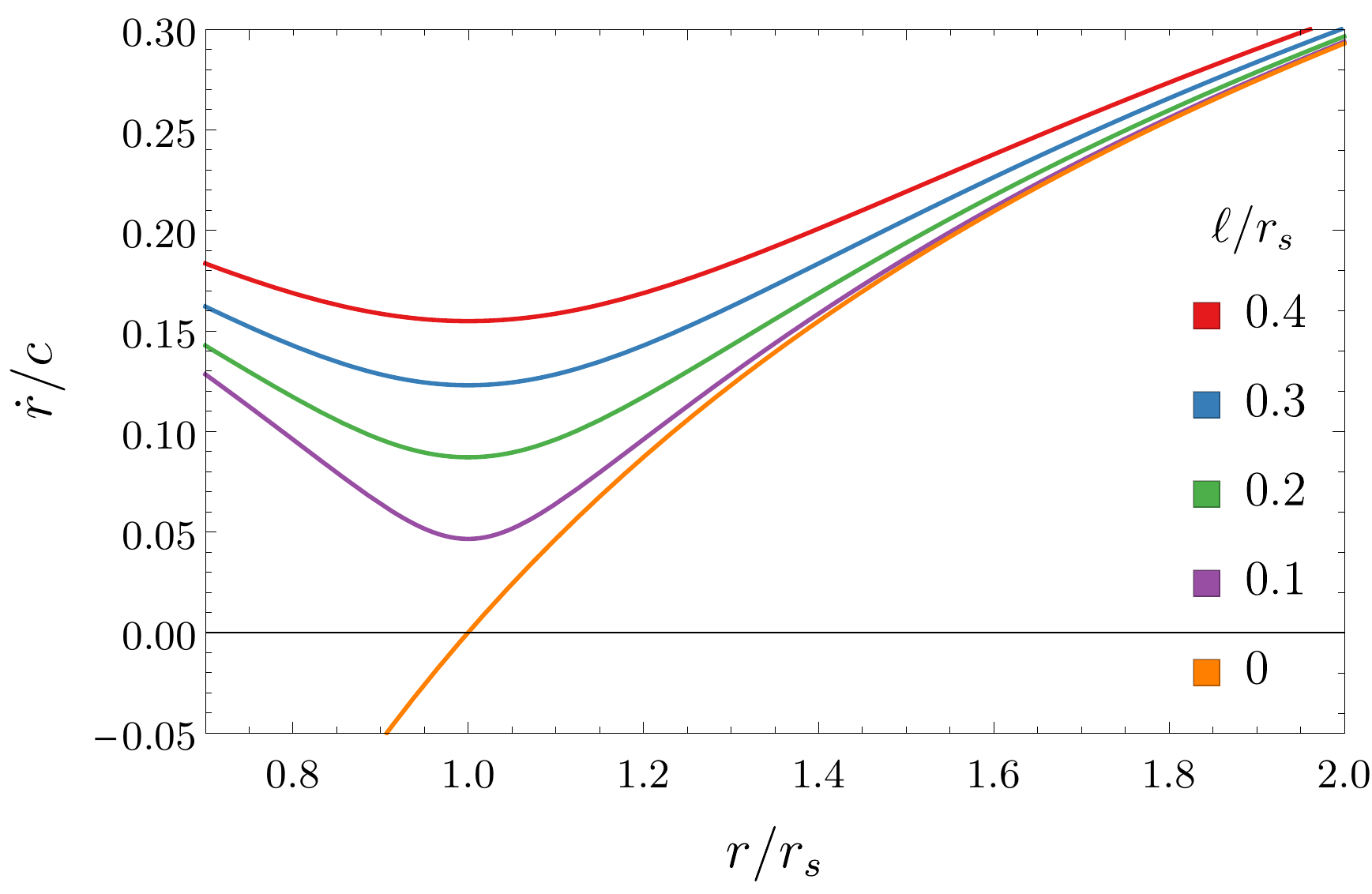}
	\includegraphics[height=50mm]{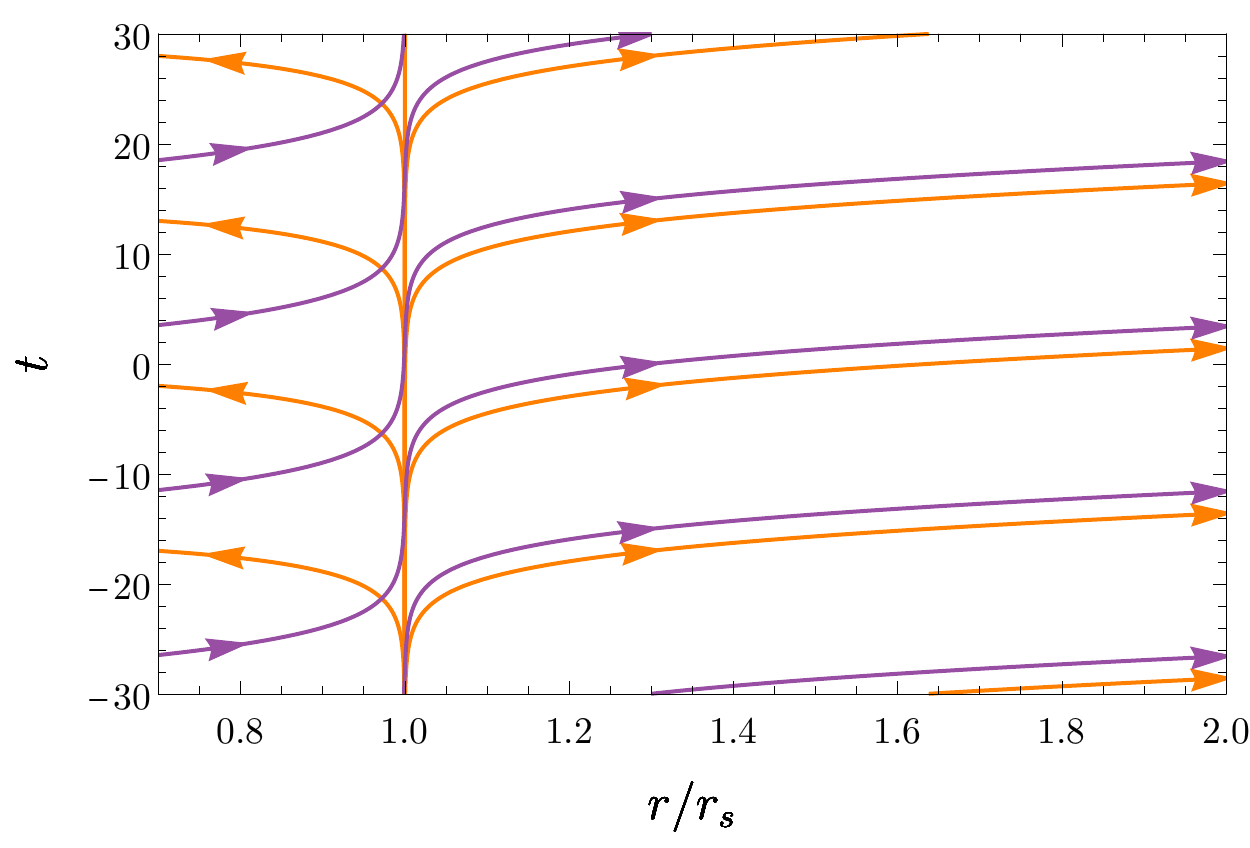}
	\caption{Velocity of light rays $\dot{r}=\dd r/\dd t$ (left) and null geodesics (right) for counter-propagating modes around the horizon for the Schwarzschild (orange) and Schwarzschild-Planck spacetimes ($\ell/r_s=10^{-3}$ on the right).}
	\label{fig.geodesics}
\end{figure}

For comparison, the velocity for null trajectories of the Schwarzschild and Schwarzschild-Planck metrics are shown in Figure \ref{fig.geodesics}. Next, we obtain the optical analogue of this regularized spacetime.

\section{Optical analogue metric} \label{sec:Analog}

\begin{figure}
	\centering
	\includegraphics[height=30mm]{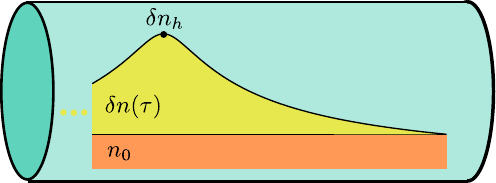}
	\caption{Pump field travelling through an optical fibre and changing the original refractive index $n_0$ due to the optical Kerr effect $\delta n(\tau)$. The pulse shape corresponds to the Schwarzschild-Planck metric.}
	\label{fibra}
\end{figure}

To obtain the optical analogue of spacetime metric, in particular one with an event horizon, an intense light signal is sent through an optical fibre. This is the pump field and it changes the refractive index of the fibre $n(\tau)$ through the optical Kerr effect $\delta n(\tau)$, this is, $n(\tau)=n_0+\delta n(\tau)$, where $n_0$ is the refractive index of the fibre itself, including material and geometric contributions. This phenomenon results in an effective metric for the movement of additional weaker light signals---even the quantum vacuum---that form the probe field. The two contributions to $n(\tau)$ are shown separately in the diagram of Figure \ref{fibra}.

The analogue event horizon occurs when the probe field interacts with the pump field reducing its phase and group velocities such that the probe is redshifted and trapped by the pump field \cite{philbin.2008,robertson.2012}. In this work we consider the fibre dispersion $n_0$ as constant by working with a single probe frequency $\omega$. We also take into account that $n_0$ contributes to the value of $n(\tau)$ in the asymptotic region $\tau\rightarrow\infty$.

The pump field travels with group velocity $u$, defining a comoving frame of reference defined by the propagation $\zeta$ and delay $\tau$ times \cite{bermudez.2016}. The transformation from the laboratory frame to the comoving frame is given by
\begin{equation}
	\zeta=\frac{z}{u}, \quad  \tau=t-\frac{z}{u},
	\label{coord}
\end{equation}
where $\zeta$ plays the role of time and $\tau$ of distance.

Philbin et al. \cite{philbin.2008} obtained the metric for a probe field travelling in a dielectric medium in the comoving frame
\begin{equation}
	g_{\mu\nu}=\frac{c^4}{u^2n^2}\begin{pmatrix}
		u^2n^2/c^2-1 &-1\\
		-1&-1
	\end{pmatrix},
\end{equation}
where we restored the $c^2$ factor such that the line element $\mathrm{d}s$ has units of length 
\begin{equation}
	\mathrm{d}s^2={u^2 n(\tau)^2}\mathrm{d}{\zeta}^2-c^2\left(\mathrm{d}{\tau}+\mathrm{d}{\zeta}\right)^2,
\end{equation}
and a conformal factor was obviated since wave propagation is well approximated by null geodesics in the eikonal limit \cite{olivera.2021}.

This is the optical analogue metric. As we saw, $n(\tau)$ can be controlled through the optical Kerr effect to define specific effective metrics. To compare this metric with the general (1+1)D spacetime metric in Eq. \eqref{met1+1}, we perform the following change of coordinates \cite{belgiorno.2011}:
\begin{equation}
	\mathrm{d}\zeta =\mathrm{d}\zeta_s-\alpha \mathrm{d}\tau, \quad \alpha=\frac{g_{01}}{g_{00}},
\end{equation}
and defining a new space coordinate $\dd \bar{\tau}=[un(\tau)/c] \dd \tau$, the metric takes the form
\begin{equation}
	\mathrm{d}s^2=\left(\frac{u^2 n(\tau)^2}{c^2}-1\right)c^2\mathrm{d}\zeta_s^2-\left(\frac{u^2  n(\tau)^2}{c^2}-1\right)^{-1}c^2\mathrm{d}\bar{\tau}^2.
	\label{metopt}
\end{equation}
From where we can directly calculate the surface gravity
\begin{equation}
	\kappa(\tau)=\left|\frac{c}{2}\frac{\dd}{\dd \bar{\tau}}\left[\frac{u^2 n(\tau)^2}{c^2}-1\right]  \right|.
	\label{kappao}
\end{equation}

As the analogue systems recreate the light trajectories at late times, we calculate the geodesics 
\begin{equation}
	\frac{\mathrm{d}\zeta}{\mathrm{d} \tau}=\left[1\pm \frac{un(\tau)}{c}\right]^{-1},
	\label{go}
\end{equation}
to connect to the astrophysical system. Particularly, the counterpropagating trajectories $(-)$ are the ones that contribute to the Hawking radiation, which is the main object of study in optical analogues \cite{philbin.2008,olivera.2021}.

\section{Optical analogue of the Schwarzschild-Planck metric} \label{sec:Spo}
In optics, it is possible to control the shape of $n(\tau)$ with the correct pump field to obtain different (1+1)D effective metrics. As we said, our goal is to establish a connection between the Schwarzschild-Planck metric in Eq. \eqref{sp} and the metric of an optical analogue in Eq. \eqref{metopt}. Matching the surface gravity in both systems given by Eqs. \eqref{kappa} and \eqref{kappao}, we obtain the following relation for the refraction index
\begin{equation}
	n(\tau)=\frac{c}{u}\sqrt{\frac{\tau_s}{\tau}},\qquad n(\tau)=\frac{c}{u}\sqrt{\frac{\tau_s}{\tau_s+ \sqrt{\left(\tau-\tau_s\right)^2+\tau_\ell^2}}}.
	\label{nsp}
\end{equation}
The same equations can be obtained by matching these metric through their geodesics in Eqs. \eqref{g} and \eqref{go}. As in the optical analogue the delay time $\tau$ takes the role of distance, the Schwarzschild radius $r_s$ and the new length $\ell$ are replaced by the time scales $\tau_s$ and $\tau_\ell$, respectively.

\begin{figure}
	\centering
	\includegraphics[height=50mm]{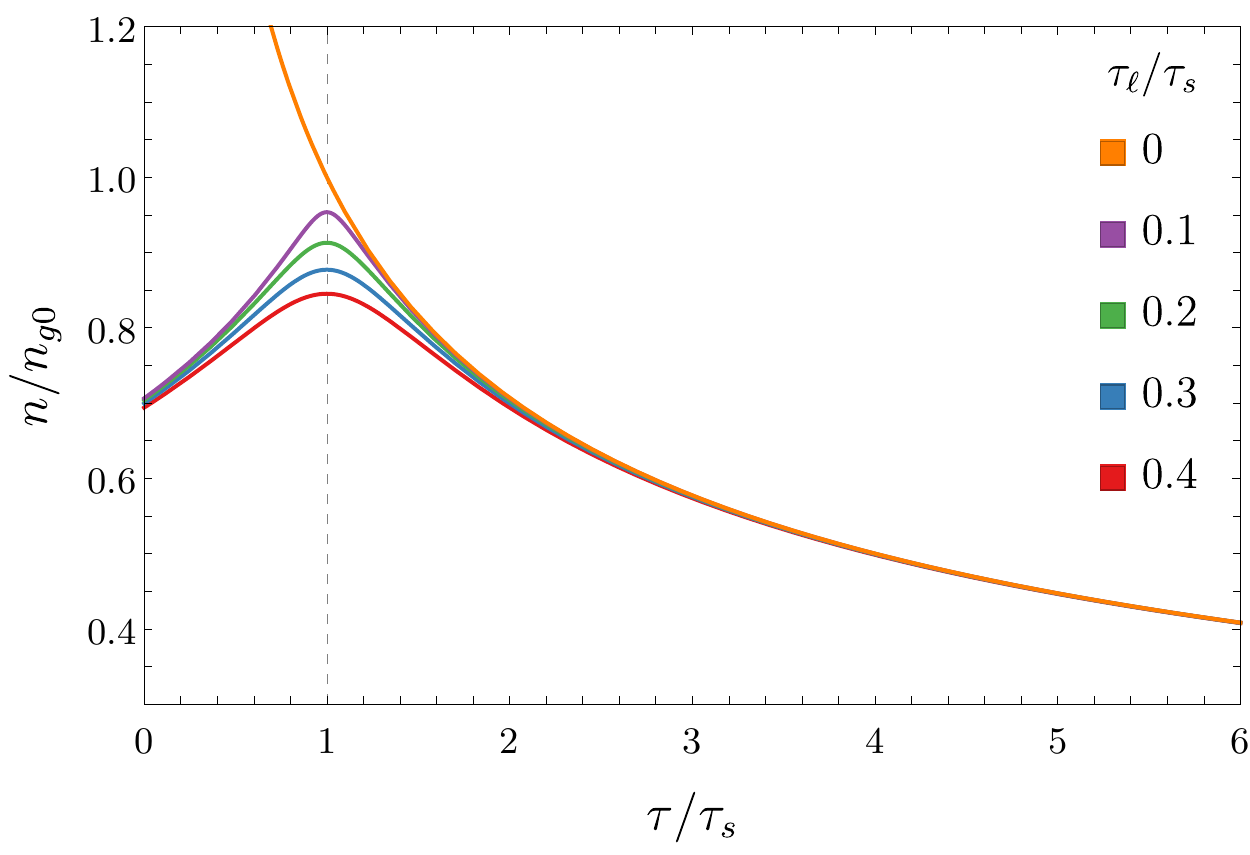}
	\includegraphics[height=50mm]{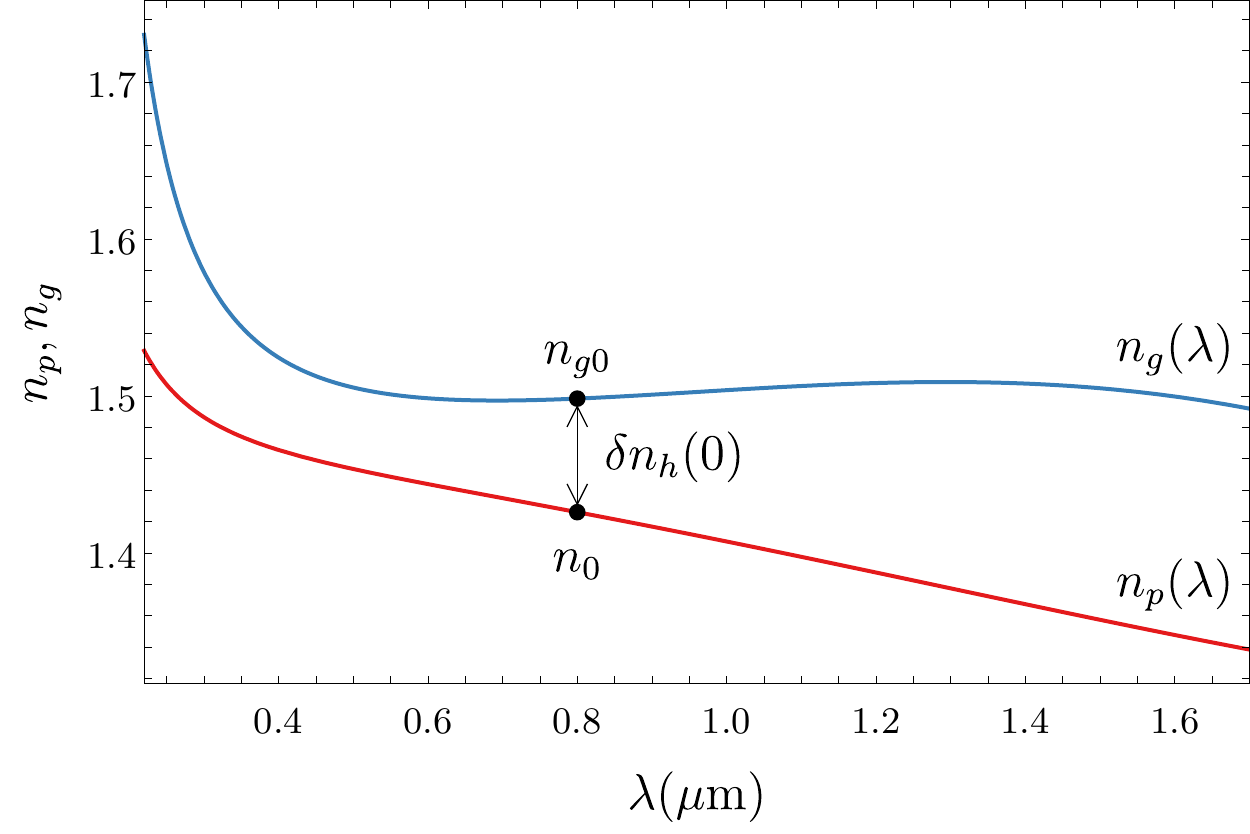}
	\caption{(a) Refractive index that recreates the Schwarzschild (orange) and Schwarzschild-Plank metrics with different values of $\tau_\ell/\tau_s$, distance is in units of horizon $\tau_s$. (b) Refractive phase index (red) and group index (blue) as function of wavelength $(\lambda)$ for a fibre used in recent optical analogue experiments \cite{Drori2019}. The dots highlight the values for a pump field at $0.8\mu\text{m}$ and $\delta n_\text{max}(0)$ is the maximum change in refractive index.}
	\label{fig.sch-p}
\end{figure}

We can control the refractive index $n(\tau)=n_0+\delta n(\tau)$ with the intensity of the pump field $I$ due to the optical Kerr effect as $\delta n(\tau)\propto I(\tau)$. The shape of the pump field that leads to the Schwarzschild spacetime is shown in Figure \ref{fig.sch-p} (the orange line). The new parameter $\tau_\ell$ smooths the pump field in the Schwarzschild-Planck metric as it is clear from the behaviour of $n(\tau)$ as the relation $\tau_\ell/\tau_s$ grows, also shown in Figure \ref{fig.sch-p}. One of these profiles is used for the diagram in Figure \ref{fibra}.

The maximum value of $\delta n$ needed to recreate the analogue Schwarzschild-Planck spacetime is reached at the horizon and is given by
\begin{equation}
	\delta n_h(\tau_\ell)=n_{g0}\sqrt{\frac{\tau_s}{\tau_s+\tau_\ell}}-n_0.
\end{equation}
The largest value among these corresponds to the limiting Schwarzschild case $\delta n_\text{max}(\tau_\ell\rightarrow 0)=n_{g0}-n_0$, with $n_{g0}=c/u$ the group velocity of the pump, as shown in Figure \ref{fig.sch-p}. This value is large but is not out of the order of possibilities, and remember this is the largest value of any possible analogue spacetimes. 

In the gravitational case, the length scale $\ell$ is related to hypothetical structure of spacetime. In the optical case $\tau_\ell$ is related to the very real microscopic structure of the fibre. We believe this parameter should be related with the response time of the fibre. For a silica fibre, the response time of the molecules to the electromagnetic field is $ \tau_\ell = 1 $ ps \cite{Boyd.2008}.

\section{Properties of the optical Schwarzschild-Planck metric} \label{sec:Properties}

\subsection{Radiation}
As we said, one of the advantages of studying accelerating mirrors is that the Bogoliubov $\beta$-coefficient can be obtained analytically \cite{Good.2016prd,Good.2020mpl}. As the Schwarzschild-Planck metric was obtained from the quantum-pure black mirror, its expression for $\beta$ is identical and given by
\begin{equation}
	\beta_{\omega \omega'}=-\frac{\sqrt{\omega \omega'}}{\pi \kappa \omega_p} e^{-\frac{\pi \omega}{2\kappa}} K_{\frac{i\omega}{\kappa}}\left(\frac{\omega_p}{g}\right)
\end{equation}
where $\omega_p=\omega+\omega'$ and $K$ is the modified Bessel function. From this quantity we can obtain the spectrum $N_\omega$ and the total number of particles $N$ as
\begin{equation}
	N_\omega=\int_{0}^{\infty}\big|\beta_{\omega \omega'} \big|^2 \dd \omega',\qquad N=\int_{0}^{\infty}\int_{0}^{\infty}\big|\beta_{\omega \omega'} \big|^2 \dd \omega'\dd \omega.
\end{equation}
The number of particles and energy are finite for the regularized metric, their values can be obtained by numerically integrating a large-enough frequency range.

\begin{figure}
	\centering
	\includegraphics[height=50mm]{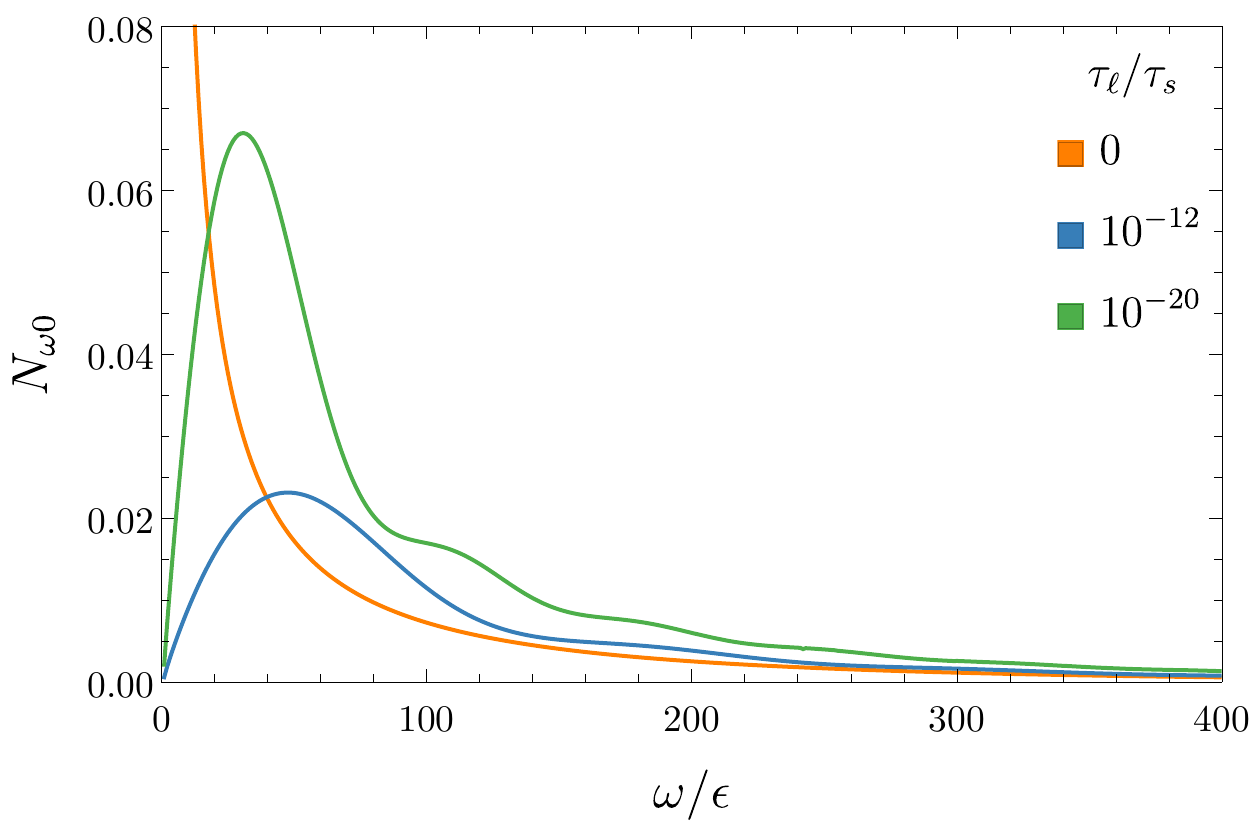}
	\includegraphics[height=50mm]{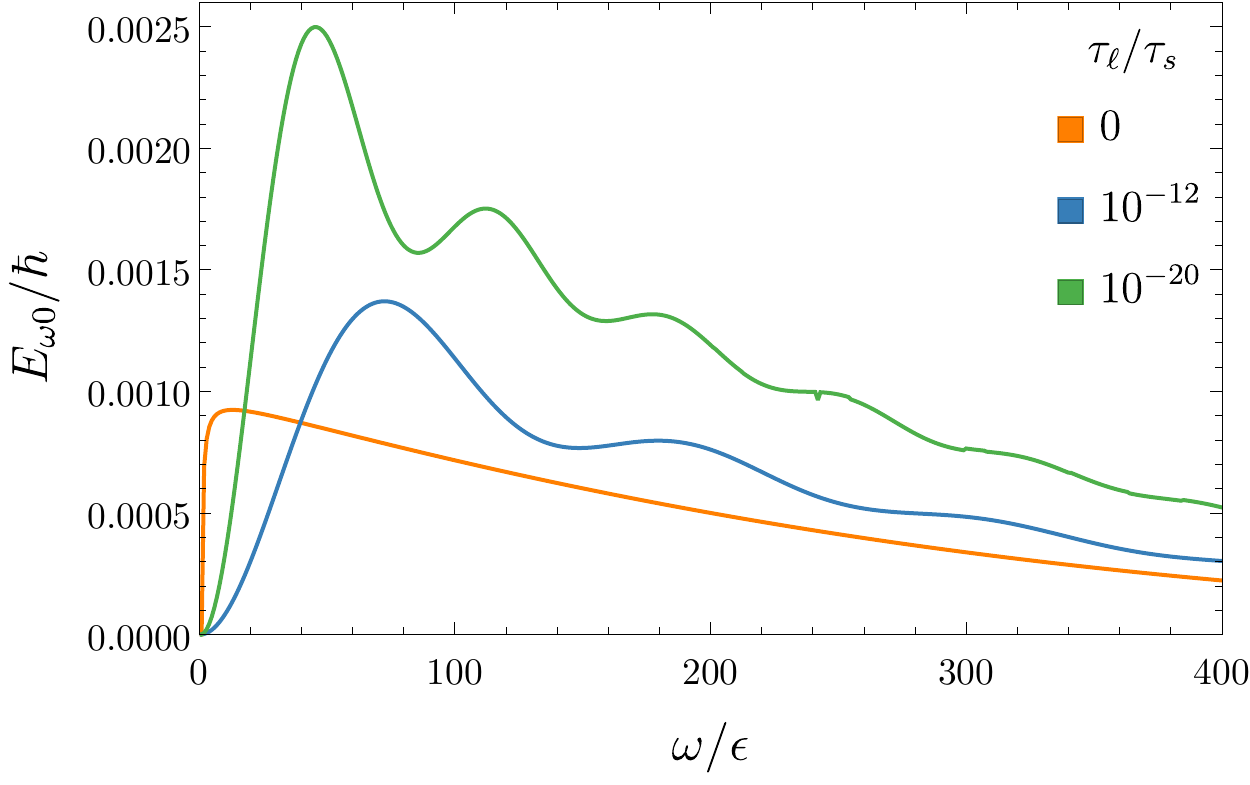}
	\caption{Discrete spectrum for the Schwarzschild $\tau_\ell/\tau_s=0$ (orange) and Schwarzschild-Planck $\tau_\ell/\tau_s=10^{-12}$ (blue), $10^{-20}$ (green) metrics with $\epsilon=0.001$. The $j$ axis is the discretized energy $\omega$ and $N_{j0}$ is the particle number in the time of collapse. The Schwarzschild spectrum is thermal as expected, and the Schwarzschild-Planck spectra are quasi-thermal, since the structure parameter $\tau_\ell$ causes deviations. }
	\label{fig.spec}
\end{figure}

To study the particle production in time and frequency, it is useful to construct a complete orthonormal family of wave packets from the Bogoliubov coefficients \cite{good.2020prd}:
\begin{equation}
	\beta_{jm\omega'}=\frac{1}{\sqrt{\epsilon}}\int_{j \epsilon}^{(j+1)\epsilon}\dd \omega e^{2 i \pi \omega m/\epsilon}\beta_{\omega \omega'},
\end{equation}
where $u=2\pi m/\epsilon$, $\omega_j=j \epsilon$, $j\geq 0$, $m$ is a natural number, and $j$ and $m$ are the discrete representations of frequency and time, respectively. The wavepacket is created during the birth of the evaporating black hole, at $m=0$. In Figure \ref{fig.spec}, we present the numerical results for three values of $\tau_\ell/\tau_s$ as $0^{-12}$, $10^{-20}$ and $0$. In this plot is clear how the thermality of the Schwarzschild metric is lost for the regularized case. 

\subsection{Geometric quantities}
In this section we study some geometric quantities that can be calculated from the optical Schwarzschild-Planck metric $g_{\mu\nu}$ given by Eq. \eqref{metopt} using the results for general (1+1)D gravity presented in Section \ref{sec:G(1+1)}.

The non-zero Christoffel symbols are
\begin{align}
	\Gamma^0_{10}&=\frac{(\tau-\tau_s)\tau_s}{2[\tau_\ell^2+(\tau-\tau_s)^2](\tau_s+\sqrt{\tau_\ell^2+(\tau-\tau_s)^2})}\approx\frac{\tau_s}{2\tau(\tau-\tau_s)}+O\left(\tau_\ell^2\right),\\
	\Gamma^1_{00}&=\frac{(\tau-\tau_s)\tau}{2\left(\tau_s+\sqrt{\tau_\ell^2+(\tau-\tau_s)^2}\right)^3}\approx\frac{(\tau-\tau_s)\tau_s}{2\tau^3}++O\left(\tau_\ell^2\right),\\
	\Gamma^1_{11}&=-\frac{(\tau-\tau_s)\tau_s}{2[\tau_\ell^2+(\tau-\tau_s)^2](\tau_s+\sqrt{\tau_\ell^2+(\tau-\tau_s)^2})}\approx-\frac{\tau_s}{2\tau(\tau-\tau_s)}+O\left(\tau_\ell^2\right).
\end{align}
These values reduce to those of the (1+1) Schwarzschild metric when the structure parameter $\tau_\ell=0$ disappears, as shown with the first term of the expansion for small $\tau_\ell$.

\begin{figure}
	\centering
	\includegraphics[height=50mm]{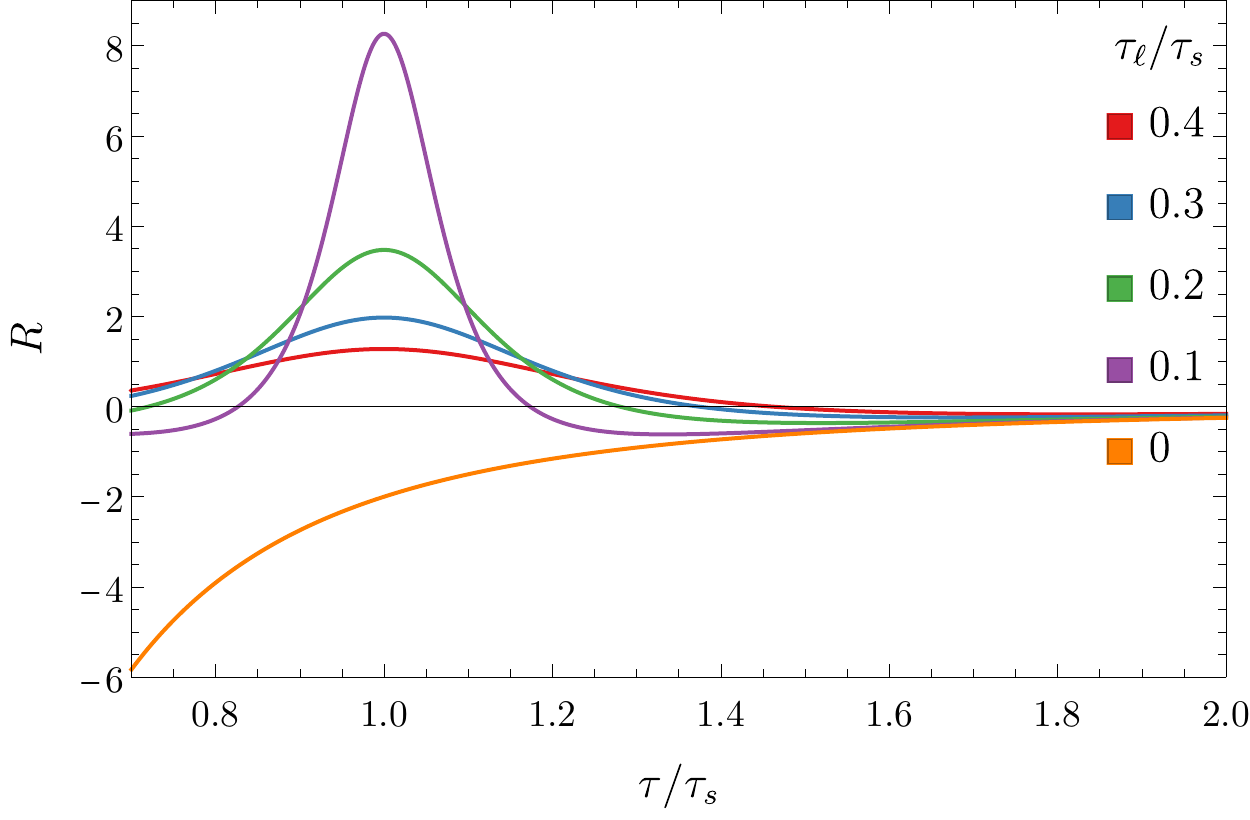}
	\includegraphics[height=50mm]{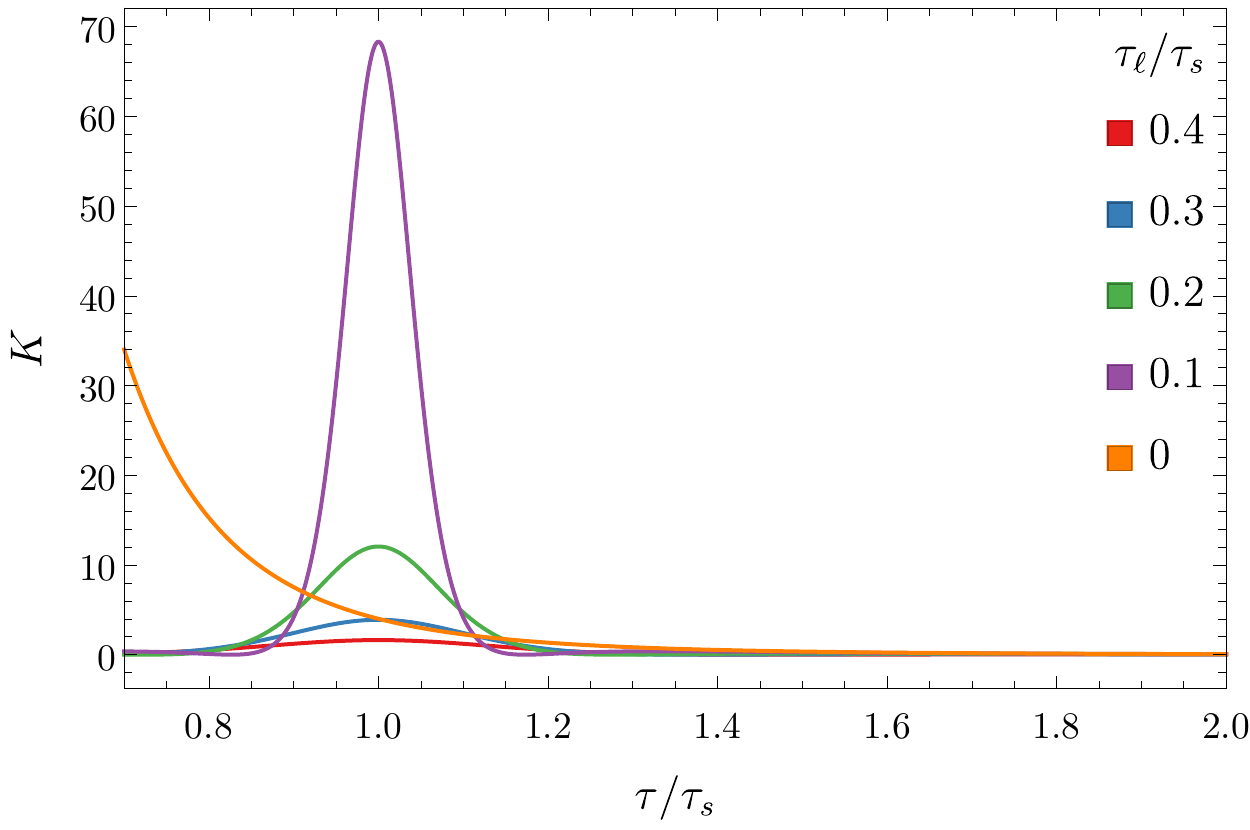}
	\caption{Ricci (left) and Kretschmann (right) scalars around the horizon $\tau_s$. The orange line is obtained from an optical analogue of the Schwarzschild metric ($\tau_\ell/\tau_\text{s}\rightarrow 0$), while the rest correspond to different cases of the Schwarzschild-Planck analogue.}
	\label{fig.scalars}
\end{figure}

The Ricci scalar is
\begin{equation}
	R=\frac{\tau_s\left[\tau_\ell^2+(\tau-\tau_s)^2\left(\tau_\ell^2-2\left(\tau-\tau_s\right)^2\right)+\tau_\ell^2\tau_s^2\sqrt{\tau_\ell^2+\left(\tau-\tau_s\right)^2}\right]}{\left[\tau_\ell^2+(\tau-\tau_s)^2\right]^2\left[\tau_s+\sqrt{\tau_\ell^2+(\tau-\tau_s)^2}\right]^3},
\end{equation}
and its limits at $\tau\rightarrow \infty$, $\tau\rightarrow \tau_s$ and $\tau_\ell \rightarrow 0$ are
\begin{equation}
	\lim_{\tau \rightarrow\infty}R=0,\quad\lim_{\tau\rightarrow\tau_s} R=\frac{\tau_s\left(\tau_\ell^2+\tau_\ell \tau_s \right)}{\tau_\ell^2(\tau_\ell+\tau_s)^3},\quad \lim_{\tau_\ell\rightarrow 0}R=\frac{2\tau_s}{\tau^3}.
\end{equation}
The Kretschmann scalar is
\begin{equation}
	K=\frac{\tau_s^2\left[\tau_\ell^2+(\tau-\tau_s)^2\left(\tau_\ell^2-2\left(\tau-\tau_s\right)^2\right)+\tau_\ell^2\tau_s^2\sqrt{\tau_\ell^2+\left(\tau-\tau_s\right)^2}\right]^2}{\left[\tau_\ell^2+(\tau-\tau_s)^2\right]^4\left[\tau_s+\sqrt{\tau_\ell^2+(\tau-\tau_s)^2}\right]^6},
\end{equation}
and the corresponding limits are
\begin{equation}
	\lim_{\tau\rightarrow\infty}K=0, \quad \lim_{\tau\rightarrow\tau_s}K=\frac{\tau_s^2\left( \tau_\ell^2 +\tau_\ell\tau_s\right)^2}{\tau_\ell^4 (\tau_\ell+\tau_s)^6}, \quad \lim_{\tau_\ell\rightarrow 0}K=\frac{4\tau_s^2}{\tau ^6}.
\end{equation}
We plot both scalars in Figure \ref{fig.scalars} around the horizon $\tau_s$, we consider the limiting case $\tau_\ell=0$ (orange line). The scalars are continuous at the horizon for all cases, but the behaviour of the Schwarzschild-Planck case differs from the Schwarzschild one around the horizon and on the black hole interior. In previous works on the quantum-pure black mirror and the astrophysical Schwarzschild-Planck spacetime, these scalars have similar expressions but are plotted as discontinuous \cite{good.2021}. We believe this is due to an implicit redefinition of the metric for the black hole interior to match the Schwarzschild case. However, the metric obtained by the optical analogue does not need this redefinition and the scalars are continuous at the horizon. The values tent to the Schwarzschild case far away from the horizon $\tau\gg\tau_s$.

Unlike the (3+1)D Schwarzschild-Planck metric, in the optical analogue the Einstein tensor is $G=0$. This result agrees with the Eq. \eqref{G}, since we are considering a spacetime (1+1)D instead of a (3+1)D one.

The non-zero entries of the Ricci tensor are
\begin{align}
	R_{00}&=\frac{\tau_s\left[\tau_\ell^4-2(\tau-\tau_s)^4-\tau_\ell^2\left(\tau^2-2\tau\tau_s-\sqrt{\tau_\ell^2+(\tau-\tau_s)^2}\tau_s+\tau_s^2\right)\right]}{2\left[\tau_\ell^2+(\tau-\tau_s)^2\right]^{3/2}\left[\tau_s+\sqrt{\tau_\ell^2+(\tau-\tau_s)^2}\right]^4}, \\
	R_{11}&=-\frac{\tau_s\left[\tau_\ell^4-2(\tau-\tau_s)^4-\tau_\ell^2\left(\tau^2-2\tau\tau_s-\sqrt{\tau_\ell^2+(\tau-\tau_s)^2}\tau_s+\tau_s^2\right)\right]}{2\left[\tau_\ell^2+(\tau-\tau_s)^2\right]^{5/2}\left[\tau_s+\sqrt{\tau_\ell^2+(\tau-\tau_s)^2}\right]^2}.
\end{align}
The non-zero entries for the Riemann tensor are
\begin{align}
	R^{0}_{110}&=\frac{\tau_s\left[ \tau_\ell^4-2(\tau-\tau_s)^4-\tau_\ell^2\left( \tau^2-2\tau\tau_s-\sqrt{\tau_\ell^2+(\tau-\tau_s)^2}\tau_s+\tau_s^2\right)   \right]}{2\left[\tau_\ell^2+(\tau-\tau_s)^2\right]^{5/2}\left[\tau_s+\sqrt{\tau_\ell^2+(\tau-\tau_s)^2}  \right]^2},\\
	R^1_{010}&=\frac{\tau_s\left[ \tau_\ell^4-2(\tau-\tau_s)^4-\tau_\ell^2\left( \tau^2-2\tau\tau_s-\sqrt{\tau_\ell^2+(\tau-\tau_s)^2}\tau_s+\tau_s^2\right)   \right]}{2\left[\tau_\ell^2+(\tau-\tau_s)^2\right]^{3/2}\left[\tau_s+\sqrt{\tau_\ell^2+(\tau-\tau_s)^2}  \right]^4},
\end{align}
while the other two can be obtained from symmetry $R^{\mu}_{\nu\beta \gamma}=R^{\mu}_{\nu\gamma\beta }$.

\section{Conclusions}\label{sec:Conclusions}

We revisited the theory that relates the trajectory of an accelerating mirror with a corresponding spacetime metric. We related these spacetime metrics from accelerated mirrors to their corresponding analogue metric. We obtained the general metric for a fibre-optical analogue of the event horizon. In particular, we constructed in two different ways the optical analogue for the Schwarzschild and Schwarzschild-Planck metrics. These metrics can be implemented in a laboratory by sending a light pulse with the corresponding shape through an optical fibre. Even more, as it is common in fibre optics, we can consider an experiment with a continuous-wave light field filling the entire length of a given fibre is more than enough. This is a good approximation during the time that the fibre includes the event horizon, given by the maximum of the pump field. The more region is included by either longer fibres or longer pump fields, the better the approximation to the analogue spacetimes and higher frequencies can be included.

The spacetime quantities obtained from the optical Schwarzschild-Planck metric in (1+1)D show a similar behaviour to that obtained by Good \cite{good.2021} for the Schwarzschild-Planck metric in (3+1)D. The only exception is the Einstein tensor, however, this is expected since in the theory of (1+1)D in the Section \ref{sec:G(1+1)} shows that the Einstein tensor does not constrain the metric in general for (1+1)D systems. This is again one of the main differences of the fibre-optical analogues of spacetime metrics and for any (1+1)D analogue systems (the majority). The consequences of this constraint should be investigated further. However, this gives us one more evidence on the robustness of the Hawking effect, as it survives even more different situations.

In this work, we mainly ignored dispersion of the fibre, since we fixed the frequency of pump wave, but this deserves further study as the dispersion also modifies the thermality of the analogue Hawking spectrum \cite{bermudez.2016,leonhardt.2012,Moreno_2020}. As it has been shown before, the dispersion can drastically change the spectrum, to a point where thermality is completely lost, but again the Hawking effect still produces particles in this situation.

Up to now, the table-top experiments of optical analogues \cite{Drori2019,philbin.2008,Rubino.2012prl} have been performed using ultra-short pulses propagating through a photonic-crystal fibre. These ultra-short pulses have a duration of $\sim$10-100 fs. If our hypothesis is correct and the microscopic time-scale of the optical metric is $\sim$1 ps, it means that these experiments would be in the regime of microscopic black-hole analogues. This would explain their high temperature emission that makes it possible to detect their analogue Hawking emission even at room temperature.

A future perspective is to use the theory we developed here to invert the process and to find trajectories of accelerating mirrors that recreate commonly used optical analogue metrics. For example, a soliton shape \cite{philbin.2008,bermudez.2016} is commonly used in the optical experiments and a tanh profile in acoustic ones \cite{robertson.2012,leonhardt.2012}.

\section*{Acknowledgments}

We thank Prof. Nora Bret\'on and Scott Robertson for their comments on an earlier version of this work. We acknowledge funding by Conacyt Ciencia de Frontera 51458-2019 (Mexico). AMR acknowledges funding by Conacyt (Mexico) scholarship 786398.

\bibliographystyle{ieeetr}
\bibliography{References}

\end{document}